\documentclass[a4paper,paper]{JHEP3}

\usepackage{graphicx}
\usepackage{epsfig,multicol,bbm}
\usepackage{psfrag}
\newcommand{\etal}{{et al.}\thinspace}
\newcommand{\rss}{\rm\scriptscriptstyle}

\title{Magnetic fields and Sunyaev-Zel'dovich effect in galaxy clusters}

\author{Rajesh Gopal\\ CAPSS, Bose Institute, \\Block - EN, Sector V, Salt Lake city, 
Kolkata - 700091, West Bengal, India\\ E-mail: \email{rgpune@gmail.com}}
\author{Suparna Roychowdhury\\ Dept. of Physics, St. Xavier's College (Autonomous),\\
30, Park Street, Kolkata - 700016, West Bengal, India\\ E-mail: \email{suparna.roychowdhury@gmail.com}}

\received{\today}
\accepted{}
\abstract{ In this work we study the contribution of magnetic fields to the Sunyaev 
Zeldovich (SZ) effect in the intracluster medium. In particular we calculate 
the SZ angular power spectrum and the central temperature decrement. The 
effect of magnetic fields is included in the hydrostatic equilibrium 
equation by splitting the Lorentz force into two terms \(\hbox{--}\) one being the 
force due to magnetic pressure 
which acts outwards and the other being magnetic tension which acts 
inwards. A perturbative approach is adopted to solve for the gas density profile for weak magnetic fields
(\( \le 4 \mathrm{\mu G}\)). This leads to an enhancement of the gas density in the central regions for 
nearly radial magnetic field configurations. Previous works had considered the force due to magnetic pressure alone which is the case only for a special set of field configurations. However, we see that there exists possible sets of configurations of ICM magnetic fields 
where the force due to magnetic tension will dominate. Subsequently, this effect is extrapolated for typical 
field strengths (\(\sim 10 \mu G\)) and scaling arguments are used to estimate the angular power 
due to secondary anisotropies at cluster scales. In particular we find that it is possible to explain the 
excess power reported by CMB experiments like CBI, BIMA, ACBAR at \(\ell > 2000\) with \(\sigma_{\rss 8} \sim 0.8\)
(WMAP 5 year data) for typical cluster magnetic fields. In addition we also see that the magnetic field effect on 
the SZ temperature decrement is more pronounced for low mass clusters (\(\langle T \rangle \sim \) 2 keV). Future 
SZ detections of low mass clusters at few arc second resolution will be able to probe this effect more precisely. 
Thus, it will be instructive to explore the implications of this model in greater detail in future works.}

\begin{document}
\section{Introduction}
It has been known that the intracluster medium has magnetic fields of micro-gauss 
strength. They affect the evolution of galaxies \cite{ars09}, contribute significantly to the 
total pressure of interstellar gas, are essential for the onset of star formation \cite{sch09}, 
and control diffusion, confinement and evolution of cosmic rays 
in the intracluster medium (ICM) \cite{kus09}. In clusters of galaxies, magnetic fields may 
play also a critical role in regulating heat conduction (e.g., \cite{chan98,nar01}), and 
may also govern and trace cluster formation and evolution. 

Magnetic fields in the intra-cluster medium have been inferred in various manners using 
diagnostics such as radio synchrotron relics within clusters, inverse Compton X-ray 
emissions from clusters, Faraday rotation measures of polarized radio sources within or behind 
clusters and cluster cold fronts in X-ray (\cite{clar01,car02}). From these observations it can 
be inferred that the medium within most clusters is magnetized with typical field strengths at a few 
\(\mu\)G level distributed throughout the cluster scale. 
In the cores of "cooling flow clusters" (\cite{eil02,tay02}) and also in cold 
fronts \cite{vi01}, the field strengths may reach 10\(\mu\)G-40\(\mu\)G and
could be dynamically important. It is thus essential to study the effect of an intra-
cluster magnetic field on the gas density distribution in the ICM. 

A very important observational probe of the cluster gas density distribution is the thermal Sunyaev
Zeldovich effect \cite{sun70,suny72}. This effect occurs because of the re-scattering of 
primary CMB photons with the hot electrons in the ICM resulting in secondary anisotropies in the 
CMB at small angular scales (for a detailed review see \cite{bir99}). At 30 GHz, anisotropies 
from the thermal SZ effect are expected to dominate over the primary CMB fluctuations for 
multipoles \(\ell \le 2500\). The angular power of SZ fluctuation depends
sensitively on the integrated cluster abundance and cluster gas distribution 
(for details see \cite{ag08}). The SZ effect in clusters depends directly on the density and 
temperature profile of the ICM. For any particular cluster, it is quantified through the y-parameter 
which is essentially the temperature decrement along a line of sight through the cluster. This is 
therefore an important observational probe and can be used to constrain/detect any additional 
parameter that affects the density of the ICM. In the present work, we study the effect of 
intracluster magnetic fields on the gas density profile of the ICM and compute the central
SZ decrement for different cluster masses as well as the angular power spectrum .  
To study magnetic field effects in the ICM, earlier authors have incorporated it by introducing
the magnetic pressure in the hydrostatic equilibrium condition (\cite{koch03,cola06, zh04}). Although 
this is an important effect, there is, in addition to the pressure, a contribution 
arising from the tension force due to the field. We show in this paper, that the most general 
treatment involves using both terms arising out of the Lorentz force due to the field. 
In general, the detailed form of the magnetic field configuration is unknown 
and moreover it can vary from cluster to cluster. It is only in the specific 
case of isotropic configurations, that the magnetic pressure is the only contribution. 
However, there could be other plausible configurations of magnetic fields in clusters where 
both the pressure and the tension force contribute. 
In this paper, we classify the different field configurations and study 
the problem for a near-radial field .  

The paper is organized as follows:
Section 2 presents the formulation of the problem by setting up the 
hydrostatic equilibrium equation incorporating the magnetic field. Sections 3 and
4 present the mathematical formulation of the y-parameter and the CMB angular power
spectrum respectively. Section 5 presents the results for the density profiles 
and SZ observables which are got by solving the hydrostatic equilibrium equation and finally  
section 6 summarizes the results and concluding remarks.

\section{Formulation}

In this section, we present the method of incorporating the effect of 
a generic tangled magnetic field for studying its effect on the ICM gas. 
We assume that the ICM is in hydrostatic equilibrium in the presence 
of magnetic field whose effect is incorporated by 
using the Lorentz force. The typically smooth morphology of the 
X-ray emission from the hot intra-cluster medium leads naturally to 
the hypothesis that the gas is in near 
equilibrium, stratified along isopotential surfaces in a mildly evolving 
distribution of dark matter, gas and galaxies. 
This suggests that the assumption of hydrostatic equilibrium for such 
relaxed clusters is mostly justified. The dark matter mass profile and 
the temperature profile are specified by their `universal' forms as 
described in the following sections. The 
radial profile of the magnetic field is also implicitly specified 
by its dependence on the 
density profile. Given these profiles, the gas density is 
finally solved perturbatively around the default profile 
(i.e the profile in the absence of magnetic field). The perturbation 
method is only applicable for small magnetic field values (in this case, 
it can be used for values of the central field upto $\sim 4\mu$G). 

\subsection{Dark matter density profile}
The gravitational potential within clusters is mainly determined by dark matter. 
The dark matter density profile, $\rho_{\rss dm}(r)$ suggested by many high 
resolution $N$ - body simulations is well described by the NFW profile 
(\cite{n96,n97})
\begin{equation}
\rho_{\rss dm} = \rho_{\rss s} y_{\rss dm}(x)=\rho_{\rss s} \frac{1}{x(1+x)^2}
\label{eq:dm}
\end{equation}
Here, $x=r/r_{\rss s}$ is the radius in unit of core radius $r_{\rss s}$ and 
$\rho_{\rss s}$ is a normalization factor which represents the characteristic 
density at , $r \, = \, r_{\rss s}$. 
Since the dark matter density profile is self-similar, the dark matter mass 
profile is also self-similar. So, the dark matter mass enclosed within a 
radius $r$ is
\begin{equation}
M(\le r) = 4 \pi \rho_{\rss s} r_{\rss s}^3 m(r/r_{\rss s})
\label{eq:mass}
\end{equation}
where, $m(x)$ is a non-dimensional mass profile given by

\begin{equation}
m(x) = \int_0^x du \,u^2 y_{\rss dm}(u) = \ln(1+x) - \frac{x}{(1+x)};
\label{eq:int}
\end{equation}

The definition of the virial radius, $R_{\rss vir}$, is the radius within which 
the total dark matter mass is confined, i.e., $M_{\rss vir}\equiv M(\le c)$, 
where
\begin{equation}
c \equiv \frac{R_{\rss vir}}{r_{\rss s}}
\label{eq:cp}
\end{equation}
is a dimensionless parameter called the 'concentration parameter'. 
Evaluating equation (\ref{eq:mass}) at the virial radius, the 
normalization factor, $\rho_{\rss s}$, is fixed at;

\begin{equation}
\rho_{\rss s} = c^3 {M_{\rss vir} \over 4 \pi R_{\rss vir}^3 m(c)}
\label{eq:rhos}
\end{equation}

The virial radius, $R_{\rss vir}\,(M_{\rss vir},z)$ is calculated with the spherical 
collapse model \cite{p80}, 

\begin{equation}
R_{\rss vir}=\left[ {M_{\rss vir} \over (4 \pi/3)\Delta_c(z)\rho_c(z)}\right] ^ {1/3} 
= \left[{M_{\rss vir}c^3 \over 4\pi \rho_{\rss s} m(c)} \right] ^ {1/3}
\label{eq:rvir}
\end{equation}
where the second equality comes from evaluating $R_{\rss vir}$ from equation 
(\ref{eq:rhos}). Here $\Delta_{\rss c}(z)$ is the spherical over-density of the 
virialized halo within $R_{\rss vir}$ at $z$, in units of the critical density 
of the universe at $z$, $\rho_{\rss c}(z)$. Following \cite{k02}, 
we assume a value $\Delta_{\rss c}(z=0) = 100$ for a cosmological model with 
$\Omega_{\rss m} = 0.29$ and $\Omega_{\rss \Lambda} = 0.71$.

We follow \cite{bull01} in adopting the approximation for $c$ as a 
function of the virial mass of the cluster. They give the median values of 
`c' and also the $1\sigma$ deviations:

\begin{equation}
c=K\Bigl ({M_{\rss vir} \over 1.5 \times 10^{13}h^{-1}M_{\rss \odot}}\Bigr )^{-0.13}
\label{eq:cpfit}
\end{equation}
with $K = 9$ reproducing the best-fit and $K = 13.5$ and $K=5.8$ reproducing 
the $+1\sigma$ and the $-1\sigma$ values in the concentration parameter.
These values of the concentration parameter are also consistent with
the findings of \cite{sel03}.
 
The above set of equations specify the dark matter density profile of a 
particular mass cluster.

\subsection{Universal temperature profile}
The ``universal temperature profile'' used for our calculation \cite{l02} 
is (normalized by the emission-weighted temperature):

\begin{equation}
{T \over \langle T \rangle} = {T_0 \over (1+{r\over a_x})^{\delta}}
\label{eq:temp}
\end{equation}
where $\langle T \rangle$ is the emission-weighted temperature of the cluster, 
$T_0 = 1.33, \, a_x = R_{\rss vir}/1.5$, and $\delta = 1.6$ on the radial range 
(0.04-1.0) $R_{\rss vir}$. To determine the emission-weighted temperature
from the cluster mass, we use a relation that arises from adiabatic evolution
of the gas in cluster. \cite{afc02} have shown that the observations 
of \cite{f01} of $M_{\rss 500} \hbox{--}\langle T \rangle$ relation 
in clusters can be understood from gravitational processes alone. We therefore 
use this empirical relation ($M_{\rss 500} \hbox{--}\langle T \rangle$) derived by 
\cite{f02}:

\begin{equation}
M_{500}=(2.64^{+0.39}_{-0.34})10^{13} \, {\rm M}_{\odot} \Bigl ( {k_b \langle 
T \rangle 
\over 1 \, 
{\rm keV}} \Bigr 
)^{1.78^{+0.10}_{-0.09}}
\label{eq:M-T}
\end{equation}
where $k_{\rss b}$ is the Boltzmann constant and $M_{\rss 500}$ has been calculated 
self-consistently by taking the total mass within the radius where the
over-density is $\delta \ge 500$. 

\subsection{Hydrostatic equilibrium of gas in presence of magnetic field}
In this section, we determine the density profile of the gas in the ICM given the universal 
temperature profile and assuming that the gas is in hydrostatic equilibrium (HE) with the 
background dark matter in the presence of a tangled magnetic field 
specified by a profile $B(r)$.

\subsubsection{Density profile in the absence of magnetic field}
The HE equation for the gas in the absence of a magnetic field has 
the well-known form:
\begin{equation}
\frac{1}{\rho_{g}(r)}\frac{dP_{g}(r)}{dr}=-\frac{GM(\leq r)}{r^{2}}
\label{eq:HEdef}
\end{equation}
where G is the gravitational constant, $\rho_{\rss g}(r)$ and $P_{\rss g}(r)$ are the density and pressure 
profiles respectively. The pressure is related to the density through the equation of state:
\begin{equation}
P(r)=\frac{\rho(r) k_B T(r)}{\mu m_p}
\label{eq:EOSdef}
\end{equation}
Here, $M(\leq r)$ is the mass enclosed within radius $r$ and $\mu$ and $m_{\rss p}$ denote 
the mean molecular weight ($\mu=0.59$) and the proton mass. $T(r)$ is 
the universal temperature profile of the gas. 
$M(\leq r)$ is mainly determined by the dark matter mass profile. The 
solution $\rho_{\rss g}(r)$ to the above HE equation is referred to as the default 
density profile.  Using the equation of state, the HE equation can be recast as:
\begin{equation}
 \frac{k_B}{\mu m_p}\left(\frac{d \ln \rho_g}{dr}+\frac{d \ln T(r)}{dr}\right)=-\frac{G M(\le r)}{r^2 T(r)}
\end{equation}
The solution $\rho_{\rss g}(r)$ to the above can then be expressed as :
\begin{equation}
\frac{\rho_g(r)}{\rho_g(0)}=\frac{T(0)}{T(r)}\exp\left(-\frac{\mu m_p}{k_B}\int_0^r 
dr \frac{G M(\le r)}{r^2 T(r)}\right)
\label{eq:dendef}
\end{equation}
The central density $\rho_{\rss g}(0)$ is fixed by the constraint that the gas mass within 
the cluster is a 
'universal' fraction of the total mass of the cluster. Thus, the gas fraction 
$f_{\rss gas}={M_{\rss gas}}/M_{\rss total}$ 
within the virial radius is taken to be universal and equal to $0.105$, as recently 
found by
Ettori (2003) for a sample of low and high redshift clusters. The gas mass can be 
expressed as:
\begin{equation}
M_{\rss {gas}}=4 \pi\int_{0}^{r_{\rss vir}} dr r^2 \rho_{\rss g}(r)
\label{eq:BCdef}
\end{equation}
Since the total gas mass is negligible compared to the dark matter, $M_{\rss total} 
\approx M_{\rss dm}$, 
mass in dark matter, and therefore $f_{\rss gas} \approx {M_{\rss gas} / M_{\rss dm}}$.

\subsubsection{Density profile in the presence of magnetic field}
The effect of magnetic fields is now taken into account by introducing an 
additional term corresponding to the radial 
component of the Lorentz force due to the field. The Lorentz force due to a 
tangled magnetic field $\vec{B}(\vec{r})$ is given by 
\begin{equation}
\vec{F}(\vec{r})=\vec{B}(\vec{r})\times(\nabla\times\vec{B}(\vec{r}))
\label{eq:Lorforce}
\end{equation} 
The Lorentz force depends on the magnetic field configuration which is not 
known and moreover its detailed form could vary from cluster to cluster.
However since we are looking at only the radial profiles, we incorporate the 
effect of magnetic field by 
considering the angle-averaged radial Lorentz force $\langle F_{\rss r} \rangle$. We then 
simplify
the angle-averaged Lorentz force by explicitly writing it down in terms of the angular
averages of the field components (i.e in terms of $\langle B_{\rss r}^2 \rangle, \langle 
B_{\rss \theta}^2 \rangle,
\langle B_{\rss \phi}^2 \rangle$).
This method has been used in the context of studying magnetic field effects on the 
spherical
collapse of non-rotating low-mass gas clouds \cite{chi94} and also in the
context of spherical accretion onto a black-hole in the presence of magnetic field 
\cite{sc83}.

It is important to note here, that the Lorentz force due to magnetic field consists
of two terms, one term which is a pure gradient \(\vec{\nabla} B^2\), known as magnetic
pressure and the other one being \((\vec{B}.\vec{\nabla})\vec{B}\) known as magnetic 
tension. Although these two terms contain the gradient operator \(\vec{\nabla}\), 
they are not identical. The magnetic pressure is a pure 
gradient \(\partial B^2/\partial r\), and hence is directly tied to the gradient of 
the underlying gas density distribution. So for instance, if the gas density profile 
is nearly flat, then magnetic pressure has a negligible contribution. On the 
other hand, the magnetic tension is not a pure gradient. The magnetic tension 
component along the radial direction can be written 
as: \((\vec{B}.\vec{\nabla}) B_r = \vec{\nabla}.(\vec{B} B_r )\), 
(using the fact that \(\vec{\nabla}.\vec{B}=0\)). Hence this tension component
is a divergence and hence consists of two sub-terms, one of which is 
like \(\partial B^2/\partial r\) (pressure-like term) and the other is like 
\(B^2 / r\). This non-pressure like term also known as "hoop stress" acts opposite
to the pressure gradient and increases towards the center. Refer to Eq A11 and 
A12. We use eq A12 for the derivation of density profiles. We have also included
the very special single case of equal field components also for reference. It is
only in this case that the Lorentz force has a pressure-like term. But in our
work we consider generic cases in which the hoop stress contributes.  

We classify the different sets of field configurations in terms of the 
relative strengths of 
the radial and transverse correlations (For details refer to the Appendix).
Previous studies have mainly focused on the isotropic configuration in which the 
relative strengths of
the three correlations are equal. In this case it can be shown, that the angle-
averaged Lorentz force
acts like a magnetic pressure. However this is only a special configuration. 
We consider a generic set of configurations in which the radial component of the 
magnetic field 
is auto-correlated , while the correlation between the 
transverse(polar) components is negligibly small.  

To begin with, the HE equation in the presence of tangled magnetic fields has the 
form:
\begin{equation}
\frac{1}{\rho_{\rss B}(r)}\frac{dP_{\rss B}(r)}{dr}=-\frac{GM(\leq r)}{r^{2}}+\langle F_{\rss r} \rangle
\label{eq:HEmag}
\end{equation} 
In this case too, the 
equation of state is specified by 
\begin{equation}
P_{\rss B}=\frac{\rho_{\rss B}(r)k_{\rss b}T_{\rss B}(r)}{\mu m_{\rss p}}
\label{eq:EOSmag}
\end{equation}
where $\rho_{\rss B}(r)$, $P_{\rss B}(r)$ and $T_{\rss B}(r)$ are the modified density, pressure and temperature
profiles respectively in the presence of the magnetic field.

In order to attempt to solve the equation for the modified density profile 
$\rho_{\rss B}(r)$ we make 
some simplifications. Firstly we assume that the field-strength is not very
large so that a perturbative solution around the default density profile $\rho_{\rss g}(r)$ 
can be attempted. For this to be true, the energy density in the magnetic field 
should 
be small compared to the gravitational energy density corresponding to the default 
gas density. 
The next thing to note is that the change in the default 
temperature profile will be negligible (i.e $T_{\rss B}(r)=T_{\rss g}(r)$) since 
we do not consider any additional energy transfer to the gas from the 
magnetic field.  We use the universal temperature profile as an input to 
deduce the density profile since its a good fit to observations. Since we
use a perturbative scheme, the change in the temperature profile due to
magnetic field contributes at only higher order in perturbation theory
and hence can be neglected. 
Under these assumptions we can combine the two equations (\ref{eq:HEdef}) and 
(\ref{eq:HEmag}) to get:
\begin{equation}
\rho_{\rss B}(r)=\rho_{\rss g}(r)\left(1+\frac{\mu m_{\rss p}}{k_{\rss B}}\int_{r}^{a} dr 
\frac{\langle F_{\rss r} \rangle (r)}{\rho_{\rss g}(r)T(r)}\right) 
\end{equation}
In this case, the gas fraction $f_{\rss gas}={M_{\rss gas}/ M_{\rss total}}$ is considered 
to be `universal' too and equal to 0.105. 
The upper limit $a$ in the above integral is determined from the relation:
\begin{equation}
M_{\rss gas}=4 \pi\int_{0}^{r_{\rss vir}} dr r^2 \rho_{\rss B}(r)
\end{equation}

In order to compute the modified profile $\rho_{\rss B}(r)$, it remains to specify the 
radial profile of the angle-averaged Lorentz force $\langle F_{\rss r}(r)\rangle$. As 
shown in the Appendix, $\langle F_{\rss r} \rangle$, in turn, can be expressed in terms of 
the angle-averages of the squared magnetic field components viz. $\langle B_{\rss r}^2 
\rangle$,
$\langle B_{\rss \theta}^2 \rangle$ and $\langle B_{\rss \phi}^2 \rangle$. Thus the form for 
$F_{\rss r}(r)$
depends on the relation between the different component profiles. 

In the present work, we compute the results for field configurations which are nearly 
radial. i.e $B^2=B_{\rss r}^2$. For such configurations, the form 
for $F_{\rss r}(r)$ is given in eq(A12). 
It thus remains to specify only the field strength profile $B(r)$ and this is done 
below.
The magnetic field profile is assumed to be of a parametric form,
\begin{equation}
B(r)=B_*\left(\frac{\rho_{\rss g}(r)}{\rho_{\rss g}(z=0)}\right)^\alpha
\end{equation}
Here, $\rho_{\rss g}(z=0)$ is the average cosmological density of the gas at $z=0$, $B_*$ 
is measured in \(\mu\)G and $\alpha$ is a parameter $\geq 2/3$. 
The above form is suggested by the rotation measure (RM)-X-ray correlation
inferred in simulations/observations. (see, e.g., \cite{car02,dola02}).
In this paper, we assume the value $\alpha=0.9$ for our calculations.(\cite{dola02}).

\section{Compton y-parameter and central SZ decrement}

The temperature decrement of CMB due to the SZ effect is directly proportional 
to the Compton parameter ($y$). For a spherically symmetric cluster, the 
Compton parameter is given by

\begin{equation}
y=2\frac{\sigma_{\rm T}}{m_{\rm e}c^2}\int_{0}^{R}p_{\rm e}(r)dl
\label{eq:y_sph_sym}
\end{equation}
\noindent where $\sigma_{\rm T}$ is the Thomson cross-section, and 
$p_{\rm e}(r) \,= \,n_{\rm e}(r)k_{\rm b}T_{\rm e}(r)$ is the electron 
pressure of the ICM,
where $n_{\rm e}(r)$ = $0.875 (\rho_{\rm gas}/m_{\rm p})$ is the 
electron number density, $k_{\rm b}$ is the Boltzmann constant, and $T_{\rm e}(r)$ is 
the electron temperature. The integral is performed along the line$\hbox{--}$of
$\hbox{--}$sight ($l$) through the cluster and the upper limit of the integral ($+R$) 
is the extent of the cluster along any particular line$\hbox{--}$of$\hbox{--}$sight.
We do not include the effects of beam size in calculating the $y$ parameter. This
approximation is justified by the fact that the pressure profiles are relatively 
flat in the inner region. The variation of pressure integrated along the line--of--sight
as a function of the projected radius is even flatter 
thus providing more justification for the above approximation.

The angular temperature profile projected on the sky due to SZ effect, 
$\Delta T(\theta)/T_{\rm CMB}$ is given in terms of the Compton 
parameter in equation~(\ref{eq:y_sph_sym})

\begin{equation}
\frac{\Delta T(\theta)}{T_{\rm CMB}}=g(x)y(\theta),
\end{equation}

\noindent
where $g(x)\equiv x$coth($x/2$)-$4$, $x\equiv h\nu/k_{\rm B}T_{\rm CMB}$, 
$T_{\rm CMB}=2.728$ \cite{f96}. In the Rayleigh-Jeans approximation, $g(x)\approx -2$.  
We only evaluate ``central'' SZ decrement from the pressure profiles of our models.
In this case, the integral in equation~(\ref{eq:y_sph_sym}) reduces to

\begin{equation}
y_{\rm 0}=2\frac{\sigma_{\rm T}}{m_{\rm e}c^2}\int_{0}^{R}p_{\rm e}(r)dr
\label{eq:y0}
\end{equation}

\noindent
In the Rayleigh-Jeans part of the CMB spectrum, the deviation from the black-body spectrum 
results in a decrement of the CMB temperature,

\begin{equation}
\Delta T_{\rm \mu w0} \approx -5.5\,y\,K
\label{eq:del_T}
\end{equation}

\noindent
We use the pressure profiles resulting from our model to calculate the
central SZ decrement in the temperature of the CMB.

\section{Angular power spectrum}

The angular two-point correlation function of the SZ temperature distribution  
in the sky is conventionally expanded into the Legendre polynomials:

\begin{equation}
\Big \langle \frac{\Delta T}{T_{\rm CMB}}({\mathbf {\hat n}})
\frac{\Delta T}{ T_{\rm CMB}}({\mathbf {\hat n}} + {\mathbf {\theta}}) \Big \rangle = 
\frac{1}{4\pi}\sum_{\rss \ell}(2\ell+1)C_{\rss \ell}P_{\rss \ell}(\cos\theta)
\end{equation}

\noindent
Since we consider discrete sources, we can write $C_{\rss \ell}= 
C_{\rss \ell}^{(P)} + C_{\rss \ell}^{(C)}$, where $C_{\rss \ell}^{(P)}$ is the contribution 
from the Poisson noise and $C_{\rss \ell}^{(C)}$ is the correlation among 
clusters (\cite{p80}, \S~41). We define the frequency independent part in the power 
spectrum as $C_{\rss \ell}^{*(P)}\,\equiv\,C_{\rss \ell}/g^{2}(x)$. The integral expression
of $C_{\rss \ell}^{*(P)}$ can be derived following \cite{col88} as

\begin{equation}
C_{\rss \ell}^{*(P)} = \int_{\rm 0}^{z_{\rss \rm dec}} dz{dV \over {dz}}
\int_{\rm M_{\rm min}}^{M_{\rm max}} dM {dn(M,z) \over {dM}}
|y_{\rss {\ell}}(M,z)|^2,
\label{eq:Cl}
\end{equation}

\noindent
where $V(z)$ is the co-moving volume and $y_{\rss {\ell}}$ is the angular Fourier 
transform of $y(\theta)$ given by

\begin{equation}
y_{\rss {\ell}}=2\pi\int y(\theta) J_{0}[(\ell +1/2)\theta]\theta d\theta ,
\end{equation}

\noindent
where $J_{0}$ is the Bessel function of the first kind of the integral
order $0$. In equation~(\ref{eq:Cl}), $z_{\rss \rm dec}$ is the redshift 
of photon decoupling and $dn/dM$ is the mass function of clusters which 
is computed in the Press-Schechter formalism \cite{pr74}. The mass function 
has been computed using the power spectrum for a $\Lambda$-CDM model with 
normalization of $\sigma_{\rss 8}=0.8$. We choose $M_{\rss \rm min}=5\times 
10^{13}M_{\odot}$ and $M_{\rss \rm max}=2\times 10^{15}M_{\odot}$ and 
integrate till redshift of $z=5$ instead of $z_{\rss \rm dec}$. This is 
done because the integral in equation~(\ref{eq:Cl}) is found to be 
insensitive to the upper limit in redshift beyond $z=4$, the reason 
being that the mass function is exponentially suppressed beyond that 
redshift in this mass range

\section{Results \& Discussion}

In this section, we investigate the effects of the magnetice field with nearly radial configuration 
on the density profiles, the central SZ decrement and on the SZ angular power spectrum. We also present 
a discussion on the other possible confiurations of the cluster magnetic field and the Lorentz force 
arising out of thes configurations.

\begin{figure}[!ht]
\centering
\psfrag{Angle averaged radial magnetic pressure/tension}{$\langle P_{mag} \rangle$/$ \langle T_{mag}\rangle $}
\begin{minipage}[t]{0.4\linewidth}
\centering
\includegraphics[scale=0.75, clip, trim=50 0 0 0]{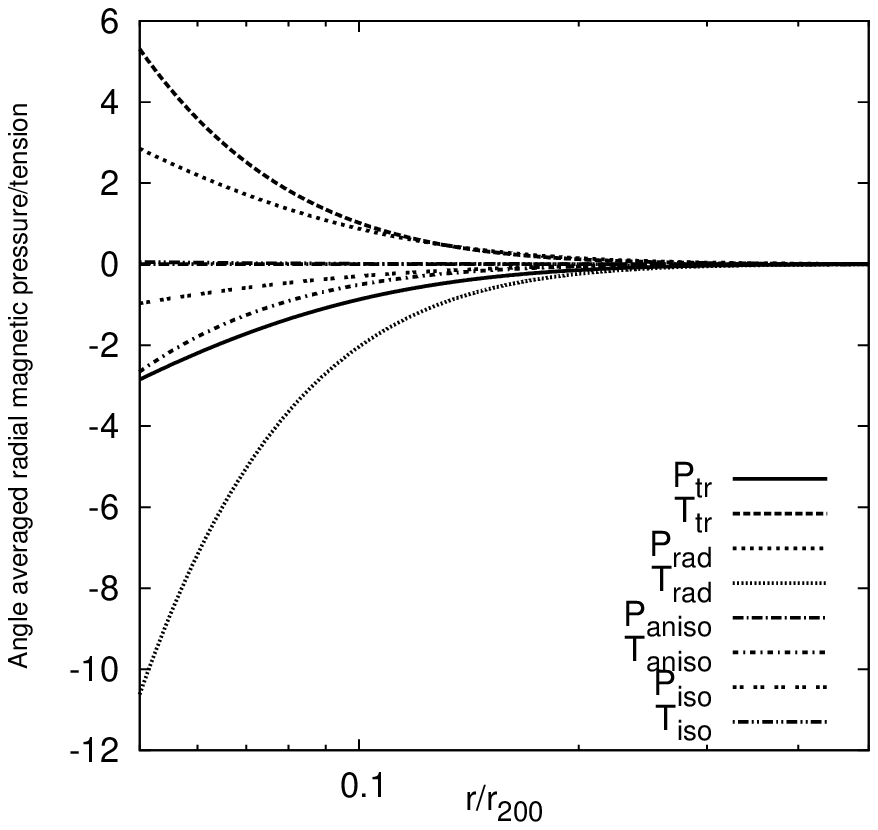}
\caption{Pressure ($P_{mag}$)and tension ($T_{mag}$) forces arising out of the magnetic Lorentz 
force as a function of the scaled distance for different magnetic field configurations like 
transverse (tr), isotropic (iso), anisotropic (aniso) and nearly radial (rad). The pressure
and tension terms are computed with arbitrary normalization}
\label{PressTension}
\end{minipage}%
\hspace{1.2cm}%
\begin{minipage}[t]{0.4\linewidth}
\centering
\includegraphics[scale=0.75, clip, trim = 52 0 0 0]{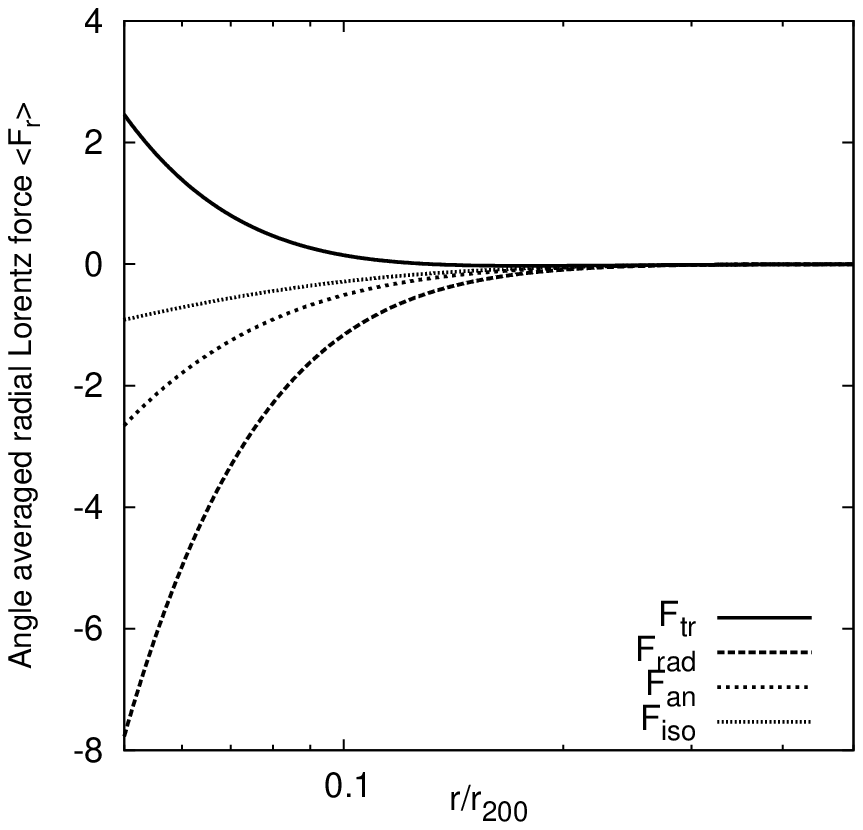}
\caption{Angle averaged Lorentz force calculated for different magnetic field 
configurations. All the nunerical factors have been normalized to
unity}
\label{Lorentzforce}
\end{minipage}
\end{figure}

\subsection{Magnetic field configurations and Lorentz force}

The Lorentz force can be split into two terms : the magnetic pressure 
($\propto\,d(B^2)/dr $) which depends on the gradient of the field 
strength and the magnetic tension ($\propto\,B^2/r$) which depends 
on the field strength (as shown in detail in the Appendix). In 
figure~(\ref{PressTension}), we plot the radial dependence of the magnetic 
pressure and the magnetic tension for different field configurations viz. radial 
(\(\langle {B_T}^2 \rangle =0\)), transverse (\(\langle {B_r}^2\rangle = 0\)), 
anisotropic (\(\langle {B_T}^2\rangle = 0.5 \langle B^2 \rangle \))and 
isotropic (\(\langle {B_r}^2\rangle = \langle{B_\theta}^2\rangle = \langle 
{B_\phi}^2\rangle = 0.33\langle B^2 \rangle \) ). In figure~(\ref{Lorentzforce}), 
we plot the resultant Lorentz force (sum of pressure and tension) for the same set of 
configurations. To demonstrate the behaviour of magnetic pressure and tension forces as well as the angle-averaged
radial Lorentz force as a function of scaled radius of the cluster, 
we have normalized all the constants to unity in both the figures. We also use the convention that positive 
values correspond to force directed radially outwards whereas negative values 
correspond to force directed radially inwards while plotting. As seen in 
figure~(\ref{PressTension}), for the \textit{radial} configuration, the tension force 
dominates over the pressure force and the net force is directed 
radially inwards as we approach the central region (\(r \le 0.2 r_{200}\)) of the 
cluster. This causes an enhancement in the gas density relative to the default 
state. This is different to the manner in which the effect of the magnetic Lorentz 
force is usually incorporated in the hydrostatic equilibrium equation for the cluster gas.

For the \textit{transverse} configuration, the tension force dominates over pressure 
force and is directed outwards as a result of which, the net force is directed 
outwards as we approach the central region. This will possibly cause a 
decrement in the gas density relative to the default case. In 
such models, the magnetic field can act as a heating source in the ICM.

In the \textit{anisotropic} case, the pressure force vanishes whereas the tension force 
is directed inwards and thus the net force is directed inwards. In the 
\textit{isotropic} configuration, the tension force vanishes whereas the pressure force is 
directed inwards and thus the net force is directed inwards. In both these 
cases, the gas tends to fall towards the central region thus causing an 
enhancement in gas density relative to the default scenario.

\FIGURE
{\epsfig{file=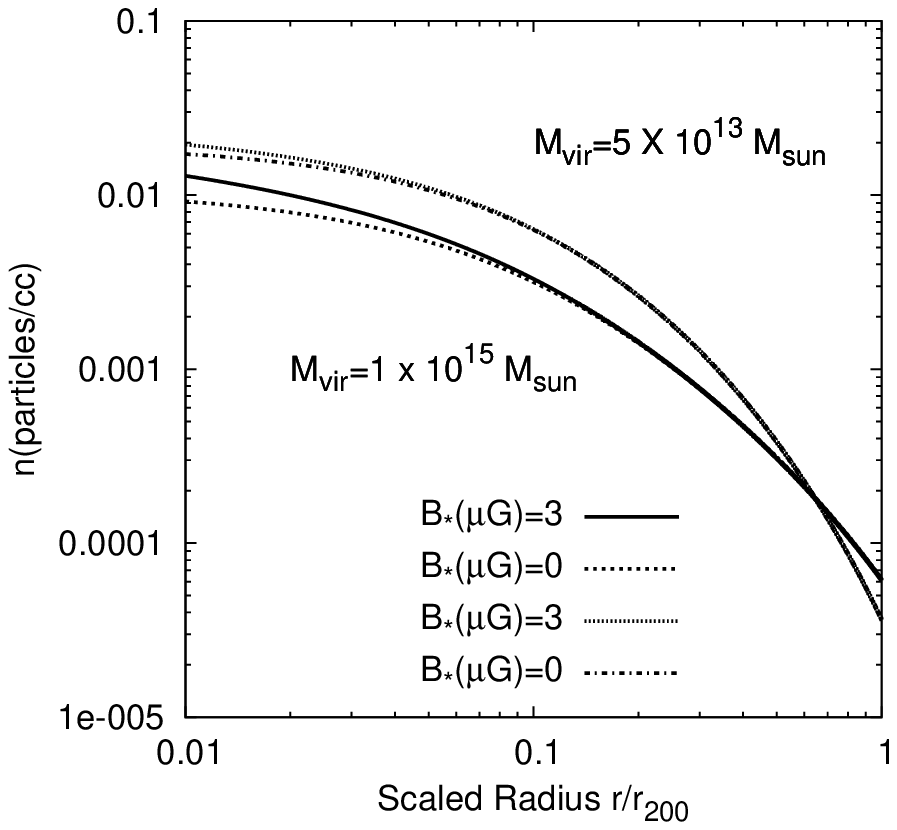,width=12cm}
\caption{Density versus scaled radius $r/r_{\rss 200}$ for  cluster masses 
$5\times 10^{13}$M$_{\odot}$ and \(1\times 10^{15}\)M\(_{\odot}\)}
\label{density}}

\DOUBLEFIGURE[!ht]{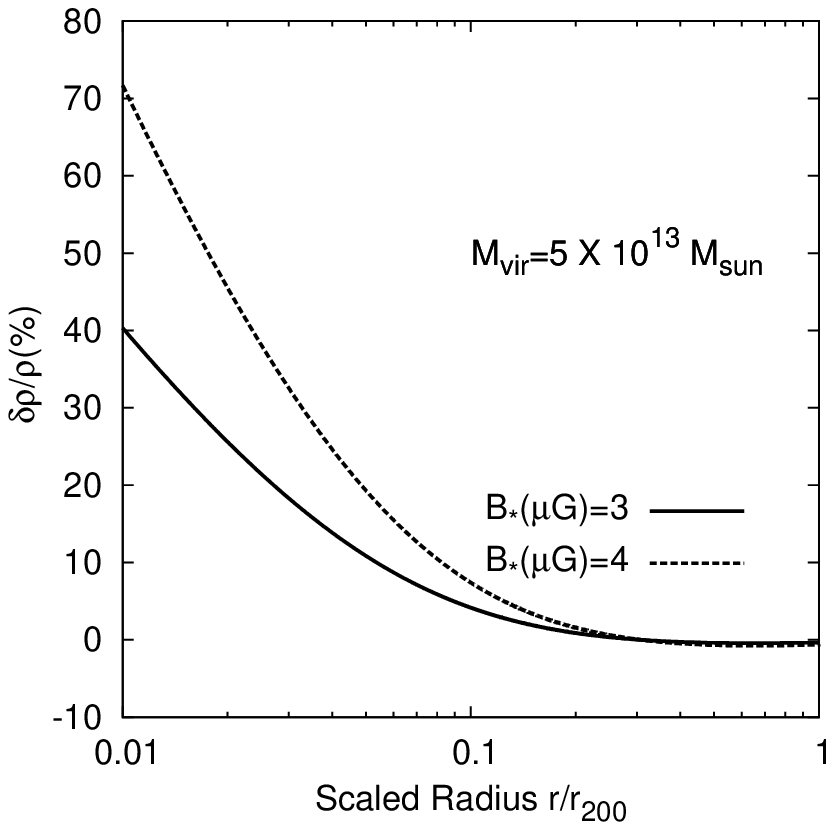,width=10cm}{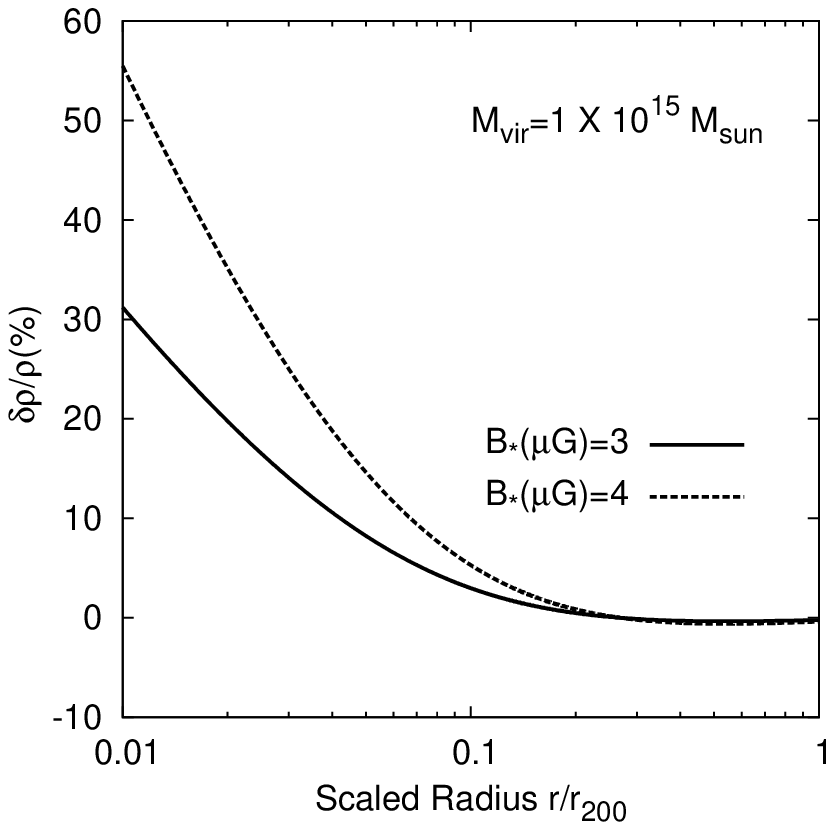,width=10cm}
{Fractional change in density due to magnetic fields of strength 3 \(\mu \)G
and 4\(\mu\)G versus scaled radius $r/r_{\rss 200}$ for a cluster mass 
$5\times 10^{13}$M$_{\odot}$}
{Fractional change in density due to magnetic fields of strength
3\(\mu\)G and 4\(\mu\)G versus scaled radius $r/r_{\rss 200}$ for a cluster 
mass $1\times 10^{15}$M$_{\odot}$}

\subsection{Density profiles}

Using the NFW profile for dark matter and the `universal' temperature profile, 
the hydrostatic equilibrium equation is solved to get the density profiles for 
different magnetic field strengths for a given cluster mass. We consider the range 
of cluster masses $5\times 10^{13}-1\times 10^{15}$M$_\odot$.

In figure~(\ref{density}), the density profile (in units of particles/cc) is 
plotted as a function of the scaled radius $r/r_{\rss 200}$ for a clusters 
of mass $5\times 10^{13}$M$_\odot$ and \(1 \times 10^{15}\)M\(_\odot\) 
for different values of field strengths specified by $B_{*}$. 
As seen in figure~(\ref{density}), the presence of magnetic field 
causes an enhancement in the gas density compared to the default(i.e no magnetic 
field) value for a particular cluster. The enhancement is large in the central 
regions and decreases outwards. There is a crossover from enhancement 
to depletion which occurs at $r_{\rss cro}=0.2 r_{\rss 200}$. 
At distances greater than the crossover radius there is a depletion in the gas 
density. This occurs because of the fact that the Lorentz force contributes through 
two opposing forces viz. the magnetic tension and magnetic 
pressure. The magnetic tension dominates in the central regions whereas magnetic 
pressure dominates in the outer regions. The magnetic tension acts inward and hence 
pushes the gas inwards increasing the density whereas in the outer 
regions(where magnetic pressure dominates) the magnetic force acts outward and 
pushes the gas outwards, depleting the density. 

In figures~4 and~5, the fractional percentage density change is plotted for 
masses $5\times 10^{13}$M$_\odot$ and $1\times 10^{15}$M$_\odot$ respectively. 
There is nearly a \(70\% \) increase in the density towards the central 
region of the cluster of mass \(5\times 10^{13}$M$_\odot\) for a 
magnetic field strength of 4\(\mu\)G. For a higher cluster mass this fractional 
change in density is lower as seen in Figure 3. The extent of density 
enhancement for a given magnetic field strength thus shows a trend with 
cluster mass, the enhancement decreasing with increasing mass. In other 
words, the effect of a magnetic field of given strength on the gas 
density is higher for low-mass clusters and groups. 
      
This enhancement of density in the central regions has important implications 
for the Sunyaev Zeldovich effect since the temperature decrement for an 
individual cluster depends directly on the gas density. This in turn 
also gets reflected in the CMB angular spectrum on small scales or 
high multipoles. We discuss these issues in the following sections.

\subsection{Central SZ decrement}
Using the density profiles, the pressure profiles are evaluated and 
the resulting central SZ decrement is computed as a function of the 
cluster mass. In figure~(\ref{CentralSZ}), the central SZ decrement 
is plotted as a function of the emission-weighted temperatures of the cluster . 
For comparison, we have also plotted the data for the central SZ 
decrement for a sample of clusters taken from \cite{zh00, mc03} and 
references therein and observed with WMAP \cite{li06}.

\FIGURE[!ht]{\psfrag{T keV}{$\langle T \rangle$ keV}
\epsfig{file=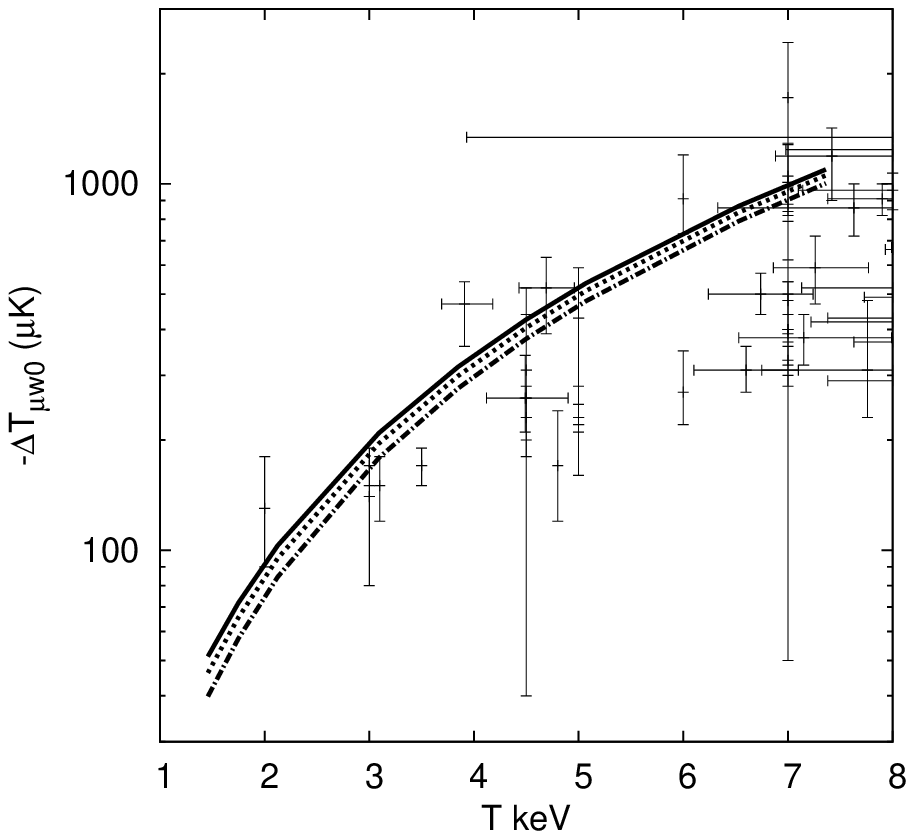,width=12cm}
\caption{Observed and predicted relations of central SZ decrement 
($\Delta T_{\mu w 0}$ ($\mu$K)) versus emission-weighted temperature 
($\langle T \rangle$ (keV)) in clusters. The solid line corresponds to 
magnetic field strength $B_{*}\,=\,4\mu G$, the dotted line corresponds to 
$B_{*}\,=\,3\mu G$ and the dot-dashed line corresponds to $B_{*}\,=\,0\mu G$. 
Data points have been taken from Zhang 
\& Wu  2000, McCarthy \etal 2003 and references therein and Lieu, Mattiz, \& 
Zhang 2006}\label{CentralSZ}}
From figure~(\ref{CentralSZ}), it can be seen that the temperature 
decrement is enhanced in the presence of magnetic fields and increases with the 
emission weighted temperature (and hence the cluster mass). The effect 
of the magnetic field is more pronounced for low mass clusters. In particular, 
magnetic fields of strength \(\sim\)3\(\mu\)G cause nearly a 25 \(\%\) 
increase in the SZ decrement for a cluster with emission weighted 
temperature \(\sim\)2 keV corresponding to a virial mass of \(1 \times 10^{14} 
M_{\odot}\). Thus from precise SZ observations of low mass clusters 
and galaxy groups it might be possible to detect or place constraints 
on the field strength in these structures.

Currently proposed SZ experiments are expected to make wide angle 
surveys of the CMB sky with a resolving power of upto few arc minutes  . 
Experiments such as South Pole Telescope (SPT) and Atacama Cosmology 
Telescope (ACT) can survey several hundreds of square degrees upto a 
mass limit 2\(\times\)10\(^{14}\) M\(_\odot\) with high sensitivity 
at arc minute resolution \cite{rr04} . In particular, the SPT survey 
will cover 4000 deg\(^2\) of sky with 10\(\mu\)K sensitivity per 1 
arc minute pixel at 150 GHz and is expected to yield nearly 10000 clusters 
with masses greater than 2\(\times 10^{14}\)M\(_\odot\) \cite{carl09}. 
The ACT survey is expected to map 200 square degrees of CMB sky in 3 
frequency bands of 145 GHz, 220 GHz and 265 GHz at arc minute resolution 
reaching sensitivity levels of 2\(\mu\)K per pixel \cite{mar09}. However 
to detect the CMB temperature decrement in the central regions it is 
necessary to be able to resolve features at the level of a few arc 
seconds. This is because the cluster virial radius (\(\sim\)1 Mpc) 
corresponds to angular scale of arc minutes whereas the central regions
(\(\sim\) 100 kpc) will typically correspond to an angular scale of few 
arc seconds. There are upcoming proposals which will probe such scales. In 
particular, the Cornell-Caltech Atacama Telescope (CCAT) \cite{gol07} 
is proposed to do high angular resolution follow-up observations
of clusters which have already been detected in some of the above mentioned 
wide-area surveys. The 150 GHz camera on CCAT would be sensitive at 
310 \(\mu\)K s\(^{1/2}\) with an angular resolution of nearly 3 arc seconds per 
pixel. These experiments will thus provide a statistically large sample 
of clusters using which it would be possible to detect or put proper 
constraints on the ICM magnetic field strengths.
	
\subsection{SZ power spectrum}
\begin{figure}[!ht]
\psfrag{l}{$\ell$}
\psfrag{lg}{$\ell(\ell + 1)\mathrm{C}_{\rm \scriptscriptstyle \ell}/2\pi \, 
(\mu K)^2$}
\psfrag{gl}{$ \Delta \mathrm{C}_{\ell}/\mathrm{C}_{\ell} (\%)$}
\centering
\begin{minipage}[t]{0.4\linewidth}
\centering
\includegraphics[scale=0.8, clip, trim = 78 0 0 0]{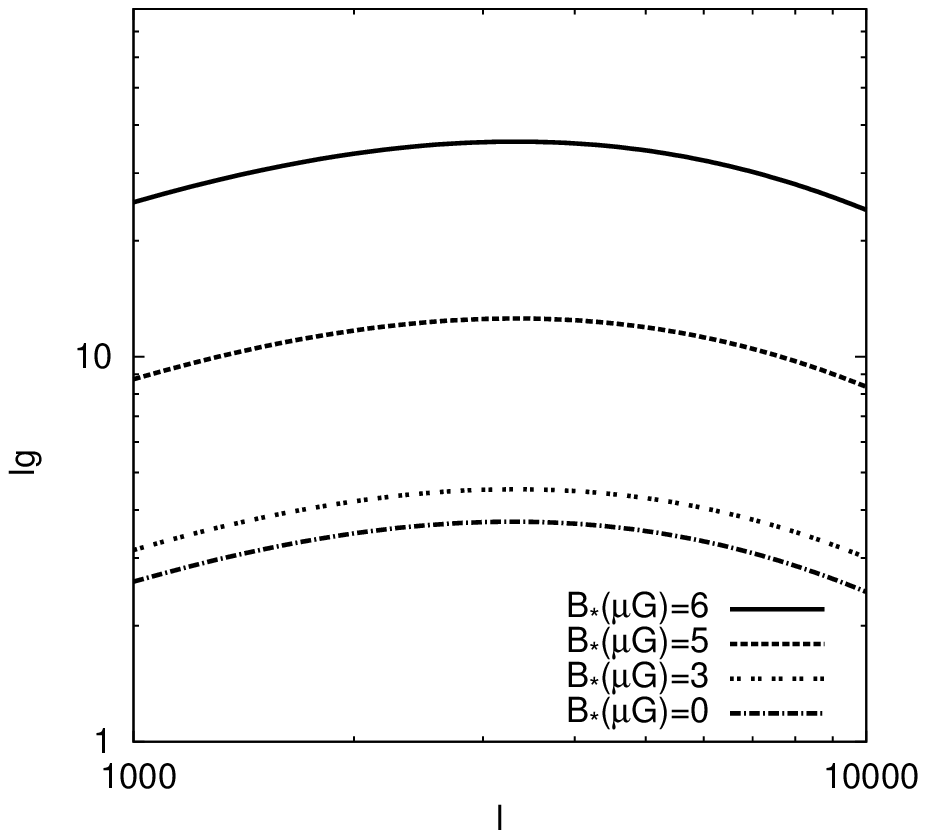}
\caption{Poisson contribution to the angular power 
spectrum of the SZ fluctuations plotted as a function of \(\ell\)}
\label{cl}
\end{minipage}%
\hspace{1.2cm}%
\begin{minipage}[t]{0.4\linewidth}
\centering
\includegraphics[scale=0.8, clip, trim = 70 0 0 0]{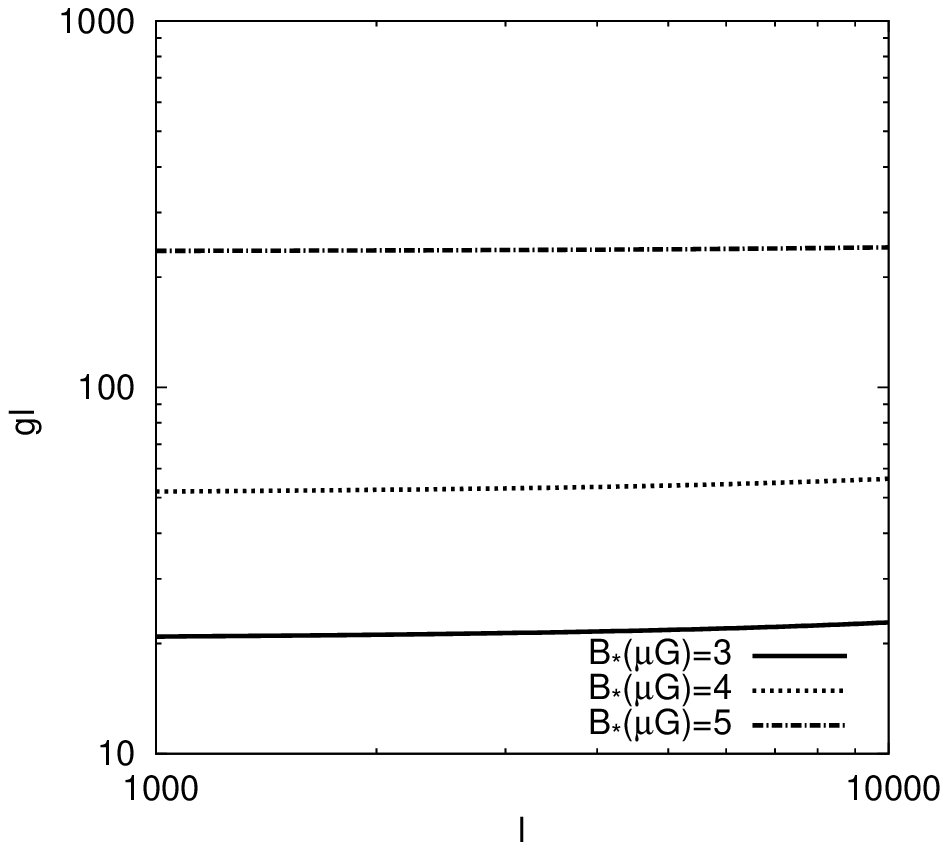}
\caption{Fractional change in angular SZ power for different magnetic 
field strengths plotted as a function of \(\ell\)}
\label{deltacl}
\end{minipage}
\end{figure}
The results for the SZ angular power spectrum are presented in figure~(\ref{cl}). 
The power spectra are plotted for different magnetic field strengths and the power spectra 
for the default density and temperature profiles is also plotted for comparison. It 
can be seen that the power peaks close to \( \ell=3300 \) for a 
field strength of \(3\mu\)G and the peak shifts to higher \(\ell\) values 
as the field strength increases. This means strong magnetic fields 
have a dominant effect on small scales. The typical value of the SZ 
power for a field strength of \(3\mu\)G is \(3.2 {\mu K}^2\). The SZ 
power increases with increasing magnetic field strength. In the figure, 
we have plotted the results for magnetic field strengths upto 
\(6 \mu\)G. The perturbative estimate is applicable only for field 
strengths upto 3\(\mu\)G. The SZ power increases with increasing magnetic 
field strength as can be seen in the figure. We can extrapolate this 
trend to estimate the CMB power at larger values of field strengths ($\sim$ 
the order of few-tens of micro-Gauss, as has been observed in clusters) by a simple 
qualitative scaling argument. Since the magnetic field depends on density as 
\(B \propto \rho^{2/3}\), and the y-parameter (which is the fractional temperature change) 
is directly proportional to density \(\Delta T/T \propto \rho\), we deduce that the 
CMB angular power which scales as temperature squared will scale with 
magnetic field as \(C_{\rss \ell} \propto B^3\). From this we can see that 
for magnetic fields of strength \(\sim\) 15-20\(\mu\)G, the CMB power 
will increase by a factor of nearly 125-300 times compared to the power 
estimated for 3 \(\mu\)G. This means that the linearly extrapolated 
estimate of CMB power for a 15\(\mu\)G field will be \(\sim\)400\({\mu K}^2\). 
In figure~(\ref{cl}), \(C_{\rss {\ell}}\) for field strengths 
5\(\mu\)G and 6\(\mu\)G are plotted using this scaling estimate.

In figure ~(\ref{cl}), we have plotted the percentage change in the angular
power of CMB compared to the default case. It can be seen from the figure
that a \(3\mu G\) field induces a 20\(\%\) increase in the CMB power whereas 
a \(4\mu G\) induces a 40 \(\%\) change.

Observations by CBI \cite{rea04} and BIMA \cite{daw06} report an excess 
in CMB anisotropy at \(\ell \ge 2000\). The primary anisotropies at 
these scales are damped out and the cause of this excess is attributed to 
the SZ power spectrum from galaxy clusters. CBI 5 year data collected at 30 GHz 
\cite{sie09} report the level of anisotropies at \(\ell \sim 3000\) to be 380 
\({\mu K}^2\) with a Gaussian error of 117 \({\mu K}^2\). An excess at nearly 
1 \(\sigma \) level is also reported by the ACBAR experiment \cite{rei09b} 
at 150 GHz at these scales. Attributing this excess to the SZ 
effect alone (without the inclusion of magnetic fields) would require the 
normalization of matter power spectrum \(\sigma_{\rss 8} \sim \) 1. 
WMAP 5 year results \cite{ko09} however have ruled out this value and fixed 
\(\sigma_{\rss 8} \sim 0.8\). As a result, this excess cannot be explained 
by the ordinary SZ effect alone and additional possible effects which might 
contribute have to be investigated (see for e.g \cite{maj08}). However as we 
discussed above, if magnetic fields are present generically at the level 
of tens of \(\mu\)G, then it might be possible to reconcile with this  
\(\sigma_{\rss 8}\sim 0.8\). 

Recent observations conducted by the Sunyaev Zeldovich Array (Sharp \etal 2009) 
do not report any excess and estimate the constraint on the level of secondary 
anisotropies to be \( 14^{+71}_{-62} {\mu K}^2 \). APEX-SZ has measured the 
CMB angular power at a frequency of 150 GHz in the range of \(3000 \le \ell \le 
10000\) and report anisotropies at the level \(33^{+37}_{-24} {\mu K}^2\) at 
effective \(\ell \sim 5000\) \cite{rei09a}. These experiments detect much 
less power compared to CBI, BIMA etc. The CMB anisotropy power at this value 
would then limit the generic magnetic field strength to \(\sim 7-8 \mu\)G. 
Several upcoming CMB experiments, such as the Atacama Cosmology Telescope(ACT), 
Planck etc claim to reach upto the cosmic variance limit and measure 
the SZ power to an accuracy of nearly 1 \(\%\) (see the science white paper 
\cite{my09}). 

The central SZ decrement and the angular power spectrum depend strongly 
on the magnetic field configuration as well. In this work we have explored
the effects of a nearly radial magnetic field. However, there could be 
other possibilities in which the effects could be different and might even be opposite
to our calculations. The extreme case where the tension force goes to zero and we are left with
only the pressure term is the isotropic configuration which has been addressed in earlier 
works by several authors \cite{cola06, koch03}. Here there is a suppression in the central
SZ decrement and thus in the angular power, since the magnetic field will act as a heating 
source and push the gas out of the central regions. In the other two cases, the effect of
the magnetic field on central SZ decrement and angular power will vary according to
the geometry of the field (as explained in the the figures~(\ref{PressTension}) and~\ref{Lorentzforce}.
Thus this can be possibly used as an indirect probe of the magnetic field configurations in the 
intracluster medium. 

Thus it is extremely important to model the effects of 
intracluster magnetic fields in addition to the standard effects in order to 
calculate the SZ power spectrum much more precisely for different field configurations.   

\section{Conclusion}
In this work, we have studied the effect of an intracluster tangled 
magnetic field on the gas distribution in the ICM. In addition, we have also 
investigated the effect of cluster magnetic fields on the central SZ 
decrement for a range of cluster masses. We have also computed the 
CMBR angular spectrum for different field strengths . In contrast to previous 
studies, we incorporated the complete radial Lorentz force due 
to the magnetic field in the hydrostatic equilibrium equations, thereby 
introducing the effect of magnetic tension as well in addition to magnetic 
pressure for a generic field configuration. It is only in the special case 
of an isotropic configuration that magnetic pressure is the sole contribution. 
However, realistic cases would involve a range of configurations. In 
particular we presented the results for a nearly radial magnetic field 
configuration. The results would be more pronounced for a configuration 
of magnetic fields in which the tension force is dominant.

It would be interesting to look at all the other important baryonic effects present in the cluster 
gas like AGN heating, cosmic-ray heating and the like in addition to this effect of the magnetic 
field to determine the structure of the intracluster gas in more 
detail and to disentangle various contributions.

Our results can thus be summarized as follows:
\begin{enumerate}
\item For a nearly radial magnetic field, the ICM gas density shows an enhancement in the 
regions close to the center whereas there is depletion in the outer 
regions, the crossover scale being dependent on the cluster mass.
\item The gas density enhancement/depletion is large for low-mass clusters.
\item The central SZ decrement is enhanced compared to the default (i.e no magnetic field) 
case if magnetic field effects are included because of the enhancement in the gas density
\item The CMB angular power is also enhanced in the presence of magnetic fields and 
future precise observations on these small scales can be used to constrain the strength 
of such fields and their configurations.
\item For an isotropic field configuration, the effects can be opposite to what we have 
concluded in this piece of work, as pointed out by earlier authors.
\end{enumerate}
  
\acknowledgments
We thank Dr. Shiv Sethi, Dr. Biman Nath and Dr. Subhabhrata Majumdar for 
discussions and suggestions.

\appendix
\section{Angle-averaged radial Lorentz force}

The Lorentz force due to a magnetic field $\vec{B}$ is given by 
\begin{equation}
{\vec{F}}=\frac{\vec{B}\times(\nabla\times \vec{B})}{4 \pi}
\label{eq:Lorentz}
\end{equation}
This can also be written as :
\begin{equation}
4 \pi \vec{F} = \frac{1}{2} \nabla (B^2)-(\vec{B}.\nabla)\vec{B}
\label{eq:Lorentz1}
\end{equation}
In spherical polar coordinates, the Lorentz force can be written in terms of its 
components as $\vec{F}=\hat{r}F_{\rss r} + \hat{\theta}F_{\rss \theta} + \hat{\phi}F_{\rss \phi}$
In particular, the radial component of the Lorentz force $F_{r}$ can be expressed as:
\begin{equation}
4 \pi F_{\rss r}=\frac{B_{\rss \phi}}{r \sin \theta}\frac{\partial B_{\rss r}}{\partial \phi}-
\frac{B_{\rss \phi}}{r}\frac{\partial}{\partial r}\left(r B_{\rss \phi} \right)-
\frac{B_{\rss \theta}}{r}\frac{\partial}{\partial r}
\left(r B_{\rss \theta} \right)+\frac{B_{\rss \theta}}{r}\frac{\partial B_{\rss r}}{\partial \theta}
\label{eq:LorentzRadial}
\end{equation}  
Adding $B_{\rss r} (\nabla.\vec{B})$ to the R.H.S this can be further rewritten 
as:
\begin{equation}
4 \pi F_{\rss r}=\nabla.(B_{\rss r} \vec{B})-\frac{{B_{\rss T}}^2}{r}-\frac{1}{2}\frac{\partial}
{\partial r}(B^2)
\label{eq:LorentzRadial1}
\end{equation}
Here, ${B_{\rss T}}^2={B_{\rss \phi}}^2+{B_{\rss \theta}}^2$, denotes the transverse part of the 
expression. We now evaluate the angular average of $F_r$. The angle averaged Lorentz force is defined as: 
\begin{equation}
\langle F_{\rss r}\rangle=\int \frac{d \Omega}{4\pi} F_{\rss r}
\label{eq:angleaveF_r}
\end{equation}
Taking the angular average of Eq~({\ref{eq:LorentzRadial1}}), we get:
\begin{equation}
4 \pi \langle F_{\rss r} \rangle = \langle \nabla.(B_{\rss r} \vec{B}) \rangle - \frac{\langle {B_{\rss T}}^2 
\rangle}{r}
-\frac{1}{2}\frac{\partial}{\partial r}(\langle B^2 \rangle)
\label{eq:AngAveFr}
\end{equation}
To get a simplified expression for the above angular average, the angle average 
$\langle \nabla.(\vec{B}B_{\rss r})\rangle$ remains to be evaluated. This can be done 
in the following manner. Let
\begin{equation}
f(r)=\langle \nabla.(\vec{B}B_{\rss r})\rangle =\int\frac{d \Omega}{4\pi}\nabla.(\vec{B}B_{\rss r})
\label{eq:fr}
\end{equation}
Multiplying both sides by $r^{2} dr$ and integrating we get,
\begin{equation}
\int dr r^{2} f(r)=\int \frac{dV}{4\pi} \nabla.(\vec{B}B_{\rss r})
\label{eq:frvol}
\end{equation}
Using Stokes theorem, the right hand side of the volume integral above can be expressed 
as a surface integral and hence,
\begin{equation}
\int dr r^{2} f(r)=r^{2} \int \frac{d\Omega}{4\pi} \hat{r}.(\vec{B}B_{\rss r})
\label{eq:Stokes}
\end{equation}
Now, differentiating both sides w.r.t $r$ we get:
\begin{equation}
f(r)=\frac{1}{r^{2}} \frac{d}{dr}(r^{2}\langle B_{\rss r}^{2}\rangle )
\label{eq:StokesDiff}
\end{equation}
Using Eq~(\ref{eq:angleaveF_r}), Eq~(\ref{eq:Stokes}) and the fact that 
$B^2=B_r^2+B_\theta^2+B_\phi^2$, the expression for the angle-averaged radial force 
can be written as 
\begin{equation}
4 \pi \langle F_{\rss r}\rangle = \frac{1}{2} \frac{d}{dr}\left[\langle B^2 \rangle - 2 \langle 
{B_{\rss T}}^2 \rangle\right]
+\frac{2 \langle B^2 \rangle - 3\langle {B_{\rss T}}^2 \rangle}{r}
\label{eq:Frave}
\end{equation}
The above expression for the angle-averaged radial Lorentz force can be simplified 
for some special magnetic field configurations which we enumerate below:
\begin{enumerate}
\item $\langle {B_{\rss \theta}}^{2}\rangle = \langle {B_{\rss \phi}}^{2}\rangle = 0$ :\\
\indent In this case, $\langle B^{2}\rangle = \langle {B_{\rss r}}^{2} \rangle$. Hence
\begin{equation}
\langle F_{\rss r}\rangle = -\frac{d}{dr}\left\langle \frac{B^{2}}{8 \pi}\right\rangle - 
\frac{4}{r}\left\langle \frac{B^{2}}{8 \pi} \right\rangle
\label{eq:Frave1}
\end{equation}
\item $\langle {B_{\rss r}}^{2}\rangle=\langle {B_{\rss \theta}}^{2}\rangle =\langle 
{B_{\rss \phi}}^{2}\rangle=\frac{1}{3}\langle B^{2}\rangle $:\\
\indent In this case, the simplified expression is,
\begin{equation}
\langle F_{\rss r} \rangle = -\frac{1}{3}\frac{d}{dr}\left\langle \frac{B^{2}}{8 
\pi}\right\rangle
\end{equation}
\end{enumerate}


\begin{thebibliography}{999}
   \bibitem{afc02} N. Afshordi, R. Cen, \apj{564}{2002}{669}
   
   \bibitem{ag08} N. Aghanim , S. Majumdar, J. Silk, 
   \newjournal{Repts.\ Prog. \ Phys.\ }{}{71}{2008}{066902}

   \bibitem{ar99} M. Arnaud , A. E. Evrard, 
   \newjournal{Mon.\ Not. \ Royal\ Astr \ Soc}{MNRAS}{305}{1999}{631}   

   \bibitem{ar05} M. Arnaud , 2005, astro-ph/0508159
   
   \bibitem{ars09} Arshakian \etal, \asas{494}{2009}{21}
   
   \bibitem{bir99} M. Birkinshaw, \prep{310}{1999}{97}
   
   \bibitem{bull01} J. S. Bullock \etal,
   \newjournal{Mon.\ Not. \ Royal\ Astr \ Soc}{MNRAS}{321}{2001}{559}

   \bibitem{car02} C. L. Carilli, G. B. Taylor, \araa{40}{2002}{319}

   \bibitem{carl09} Carlstrom \etal, 2009,  arxiv:0907.4445 

   \bibitem{ct07} A. Cattaneo, R. Teyssier,
   \newjournal{Mon.\ Not. \ Royal\ Astr \ Soc}{MNRAS}{376}{2007}{1547}
   
   \bibitem{chan98} B. D. G Chandran, S. C. Cowley , \prl{80}{1998}{3077}

   \bibitem{chi94} T. Chiueh, J-K Chou , \apj{431}{1994}{380}

   \bibitem{cbkbf01} Churazov \etal, \apj{554}{2001}{261}
   
   \bibitem{clar01} T. E. Clarke , P. P. Kronberg, H. B\"ohringer , \apj{547}{2001}{L111} 

   \bibitem{cola06} S. Colafrancesco, F. Giordano \asas{454}{2006}{L131}
   
   \bibitem{col88} S. Cole, N. Kaiser, 
   \newjournal{Mon.\ Not. \ Royal\ Astr \ Soc}{MNRAS}{233}{1988}{637}
   
   \bibitem{dalla04} Dalla Vecchia \etal,
   \newjournal{Mon.\ Not. \ Royal\ Astr \ Soc}{MNRAS}{355}{2004}{995}
   
    \bibitem{daw06} K. S. Dawson \etal, \apj{647}{2006}{13}
    
   \bibitem{don06} Donahue \etal, \apj{643}{2006}{730} 

   \bibitem{dola02} K. Dolag , M. Bartelmann, H. Lesch , \asas{387}{2002}{383} 

   \bibitem{eil02} J. A. Eilek , F. N. Owen, 2002,\apj{567}{2002}{202} 

   \bibitem{ett03} S. Ettori,
   \newjournal{Mon.\ Not. \ Royal\ Astr \ Soc}{MNRAS}{344}{2003}{L13}

   \bibitem{f01} A. Finoguenov , T. H. Reiprich , H. B\"ohringer, \asas{368}{2001}{749}

   \bibitem{f02} A. Finoguenov , C. Jones, H. B\"ohringer , T. J. Ponman, \apj{578}{2002}{74}
   
   \bibitem{f96} Fixsen \etal, \apj{473}{1996}{576} 

   \bibitem{gol07} Golwala \etal, \baas{39}{2007}{886}


   \bibitem{h00} S. F. Helsdon, T. J. Ponman , 
   \newjournal{Mon.\ Not. \ Royal\ Astr \ Soc}{MNRAS}{315}{2000}{356} 

   \bibitem{koch03} P. M. Koch , Puy D. Jetzer Ph, , 2003, 
   \newjournal{New. \ Astr. }{NA}{8}{2003}{1}

   \bibitem{k02} E. Komatsu , U. Seljak ,
   \newjournal{Mon.\ Not. \ Royal\ Astr \ Soc}{MNRAS}{336}{2002}{1256}
   
   \bibitem{ko09} E. Komatsu \etal,
   \newjournal{Ap\ Jour. \ Supp}{ApJS}{180}{2009}{330}

    \bibitem{kos06} A. Kosowsky A., 2006, New Astronomy Reviews, 50, 969
	\newjournal{New.\ Astr \ Revs}{NAR}{50}{2006}{969}
	
    \bibitem{kus09} D. Kushnir, B. Katz, E. Waxman,
   \newjournal{Jour \ Cosm. \ AstroPart.}{JCAP}{9}{2009}{24}
   
   \bibitem{li06} R. Lieu , J. P. D. Mittaz, S. N. Zhang, \apj{648}{2004}{176} 

   \bibitem{l02} C. Loken , M. L. Norman , E. Nelson , G. L. Bryan, P. Motl , \apj{579}{2002}{571}
   
   \bibitem{maj08} S. Majumdar S.  2008, J.Phys : Conf. Ser. 140 01200
    \bibitem{m98} M. Markevitch, \apj{504}{1998}{27}

    \bibitem{mar09} T. Marriage,  Atacama Cosmology Telescope Team, \baas{41}{2009}{755}

    \bibitem{mathews04} Mathews \etal, 2004, \apj{615}{2004}{662}
    
    \bibitem{mc03} I. G. McCarthy , A. Babul , M. L. Balogh, G. P. Holder, \apj{591}{2003}{526} 
    
    \bibitem{mush03} Mushotzky R. F., 2003, AIP Conference Proceedings, 
   666, 171 
   
    \bibitem{my09} Myers \etal 2009, Astro2010: The Astronomy and Astrophysics Decadal 
   Survey, Science White Papers, no. 218   
   
    \bibitem{nar01} R. Narayan, M. V. Medvedev , \apj{562}{2001}{L129}
	
   \bibitem{n96} J. F. Navarro , C. S. Frenk, S. D. M. White, \apj{462}{1996}{563}

   \bibitem{n97} J. F. Navarro , C. S. Frenk, S. D. M. White, \apj{490}{1997}{493}
   
   \bibitem{p80} P. J. E. Peebles, {\it The Large Scale 
   Structure of the Universe}, Princeton Univ. Press 1980, Princeton, NJ

   \bibitem{piff05} R. Piffaretti, P. Jetzer , J. Kaastra, T.Tamura, \asas{433}{2005}{101}

   \bibitem{p99} Ponman \etal, \nature{397}{1999}{135} 

   \bibitem{p03} T. J. Ponman, A. J. R. Sanderson , A. Finoguenov,
	\newjournal{Mon.\ Not. \ Royal\ Astr \ Soc}{MNRAS}{343}{2003}{331}

   \bibitem{pratt03} G. Pratt G., M. Arnaud, \asas{408}{2003}{1} 

   \bibitem{pratt06} G. W. Pratt, M. Arnaud, E. Pointecouteau, \asas{446}{2006}{429} 
   
   \bibitem{pr74} W. H. Press, P. Schechter \apj{187}{1974}{425}
   
   \bibitem{rea04} A. C. S. Readhead \etal \apj{609}{2004}{498}
   
   \bibitem{rei09a} C. L. Reichardt \etal \apj{701}{2009}{1958}
   
   \bibitem{rei09b} C. L. Reichardt \etal \apj{694}{2009b}{1200}

   \bibitem{rrnb04} Roychowdhury \etal,
	\newjournal{Mon.\ Not. \ Royal\ Astr \ Soc}{MNRAS}{615}{2004}{681}
	
   \bibitem{rr04} Ruhl \etal, 2004,
   \newjournal{Proc.\ SPIE.}{PrSPIE}{5498}{2004}{11}
   
   \bibitem{s03} A. J. R. Sanderson, T. J. Ponman, A. Finoguenov,
   E. J. Lloyd-Davies , M. Markevitch,
   \newjournal{Mon.\ Not. \ Royal\ Astr \ Soc}{MNRAS}{340}{2003}{989}
   
   \bibitem{sc83} E. T. Scharlemann , \apj{272}{1983}{279} 
   
   \bibitem{sch09} Schleicher \etal 2009, \apj{703}{2009}{1096} 
  
   \bibitem{sel03} U. Seljak, K. M. Huffenberger,
   \newjournal{Mon.\ Not. \ Royal\ Astr \ Soc}{MNRAS}{340}{2003}{1199}

   \bibitem{sie09} J. L. Sievers \etal , \astroph{0901.4540}
   
   \bibitem{sun70} Sunyaev R. \& Zeldovich Y. B., 1970, 
   \newjournal{Comm.\ Astr. \ SpPhys.}{CASP}{2}{1970}{66}
   
   \bibitem{suny72} Sunyaev R. \& Zeldovich Y. B., 1972, 
   Comments on Astrophysics and Space Physics, 4, 173
   \newjournal{Comm.\ Astr. \ SpPhys}{CASP}{4}{1972}{173}
   
   \bibitem{tay02} G. B. Taylor, A. C. Fabian, S. W. Allen,
	\newjournal{Mon.\ Not. \ Royal\ Astr \ Soc}{MNRAS}{334}{2002}{769}

   \bibitem{vi01} A. Vikhlinin , M. Markevitch , S.S. Murray, \apj{549}{2001}{L47}
   
    \bibitem{voit05} M. G. Voit, \rmp{77}{2005}{207}
   
   \bibitem{zh04} P. Zhang ,
   \newjournal{Mon.\ Not. \ Royal\ Astr \ Soc}{MNRAS}{348}{2004}{1348}   
   
   \bibitem{zh00} Y. Zhang, X. Wu , \apj{545}{2000}{141} 

\end{thebibliography}
\end{document}